\documentclass[conference]{IEEEtran}
\IEEEoverridecommandlockouts
\usepackage{cite}
\usepackage{amsmath,amssymb,amsfonts}
\usepackage{algorithmic}
\usepackage{graphicx}
\usepackage{textcomp}
\usepackage{xcolor}
\usepackage{balance}
\usepackage{hyperref}
\usepackage{todonotes}
\usepackage{booktabs}
\usepackage{siunitx}
\usepackage{multirow}
\usepackage{subcaption}
\usepackage{tcolorbox}

\def\BibTeX{{\rm B\kern-.05em{\sc i\kern-.025em b}\kern-.08em
    T\kern-.1667em\lower.7ex\hbox{E}\kern-.125emX}}
    
\begin{document}
\title{Visual Integration of \\Static and Dynamic Software Analysis in \\Code Reviews via Software City Visualization}

%\author{
%	\IEEEauthorblockN{Alexander Krause-Glau}
%	\IEEEauthorblockA{\textit{Department of Computer Science} \\
%	\textit{Kiel University}\\
%	Kiel, Germany \\
%	alexander@krause-glau.de}
%	\and
%	\IEEEauthorblockN{Lukas Damerau}
%	\IEEEauthorblockA{\textit{Department of Computer Science} \\
%	\textit{Kiel University}\\
%	Kiel, Germany \\
%	lukas.damerau@email.uni-kiel.de}
%	\and
%	\IEEEauthorblockN{Malte Hansen}
%	\IEEEauthorblockA{\textit{Department of Computer Science} \\
%	\textit{Kiel University}\\
%	Kiel, Germany \\
%	malte.hansen@email.uni-kiel.de}
%	\and
%	\IEEEauthorblockN{Wilhelm Hasselbring}
%	\IEEEauthorblockA{\textit{Department of Computer Science} \\
%	\textit{Kiel University}\\
%	Kiel, Germany \\
%	hasselbring@email.uni-kiel.de}
%}

\author{
	\IEEEauthorblockN{
		Alexander Krause-Glau\IEEEauthorrefmark{1},
		Lukas Damerau\IEEEauthorrefmark{2}, 
		Malte Hansen\IEEEauthorrefmark{2} and
		Wilhelm Hasselbring\IEEEauthorrefmark{3}
		}
	\IEEEauthorblockA{
		Department of Computer Science,
	Kiel University\\
		Kiel, Germany\\
		Email: \IEEEauthorrefmark{1}alexander@krause-glau.de,
		\IEEEauthorrefmark{2}\{forename\}.\{surname\}@email.uni-kiel.de
		\IEEEauthorrefmark{3}hasselbring@email.uni-kiel.de
	}
}

\maketitle

% SV sinnvoll für PC, daher sinnvoll in Code Review. Hier kommt es aber auf die einfache und nahtlose Integration in den Workflow. Typisch sind externe tools die disliked context switches erzielendie und die statisch nutzen.

\begin{abstract}
Software visualization approaches for code reviews are often implemented as standalone applications, which use static code analysis.
The goal is to visualize the structural changes introduced by a pull / merge request to facilitate the review process.
In this way, for example, structural changes that hinder code evolution can be more easily identified, but understanding the changed program behavior is still mainly done by reading the code.
For software visualization to be successful in code review, tools must be provided that go beyond an alternative representation of code changes and integrate well into the developers' daily workflow.

In this paper, we report on the novel and in-progress design and implementation of a web-based approach capable of combining static and dynamic analysis data in software city visualizations.
Our architectural tool design incorporates modern web technologies such as the integration into common Git hosting services.
As a result, code reviewers can explore how the modified software evolves and execute its use cases, which is especially helpful for distributed software systems.
In this context, developers can be directly linked from the Git hosting service's issue tracking system to the corresponding software city visualization.
This approach eliminates the recurring action of manual data collection and setup.
We implement our design by extending the web-based software visualization tool ExplorViz.
We invite other researchers to extend our open source software and jointly research this approach.
Video URL: \url{https://youtu.be/DYxijdCEdrY}
\end{abstract}


\vspace*{-0.2cm}
\section{Introduction}
% Code reviews common in SE
% change-based most common
% Problem with text-based code reviews
% SoftVis to the rescue
% Our Contribution with focus on dynamic analysis and integration into common workflow, instead of new tool with new workflow (also tool-information misalignment paper)
Code reviewing has a long history of research~\cite{badampudi2023CodeReviewLiteratureSurvey, basili1987CodeReading, fagan1976CodeInspections, soderberg2022CodeReviewMisalignments} and is nowadays a standard activity in professional software development.
It is an effective method for finding non-functional defects, such as those that affect code evolution~\cite{bachelli2013CodeReview,mantyla2009CodeReviewDefectTypes} in upcoming changes to the code base.
With a focus on the change-based style~\cite{baum2017}, developers spend an average of almost six and a half hours a week in reviewing~\cite{bosu2013CodeReviewTimeSpend}.
As software becomes more complex and may be generated by generative AI such as GitHub Copilot, the time required for code reviews can be expected to increase.
Researchers are therefore investigating how this task can be facilitated.

A promising approach is the use of software visualization (SV) in the reviewing process.
The reason for this is that SV can facilitate the underlying task of program comprehension (PC)~\cite{cornelissen2011ControlledExperimentProgramComprehension, wettel2011ControlledExperiment}.
To this end, SV approaches for PC commonly use static code analysis to visualize structural properties of the code base.
In the context of code review, for example, this means that structural code changes are visualized differently and therefore represent an alternative or supplement to the usual textual representation.
However, additional presentations of the changes introduced by a pull / merge request (from now on called change request or CR) are not sufficient for reviewers.
In fact, discrepancies have been observed between code review tools and functional defect detection~\cite{czerwonka2015CodeReviewsNoFunctionalBugs}, as well as between tooling and the information needed for the review process~\cite{soderberg2022CodeReviewMisalignments}.
Therefore, developers tend to use their familiar development environment for PC instead of using specific tools such as SV~\cite{xia2018}.
Overall, we expect that the usefulness of SV in code reviews depends, among other things, on the ability to visualize potential behavioral changes at runtime, as well as on the ease of integration into the developers' daily workflow.
As there exist already well-established procedures and tools for code review~\cite{badampudi2023CodeReviewLiteratureSurvey}, SV approaches should expand on these procedures rather than introducing a completely new review process in a standalone tool.
%Additionally, the relative position in which changed files are presented during the code review task has an impact on the review's outcome~\cite{baum2017CodeReviewOptimalOrder,fregnan2022}.

In this paper, we report on the novel and in-progress design and implementation of a web-based approach capable of combining static and dynamic analysis data in software city visualizations.
To achieve that, we extend ExplorViz~\cite{VISSOFT2013,fittkau2017IST,hasselbring2020}, which so far only used dynamic analysis data.
The result is a visualization of the changes of a software systems' structural evolution and dynamic runtime behavior.
Our web-based approach is designed to supplement the process of code review in common Git hosting services (GHS).
In this way, reviewers are able to explore a software system's visualized changed runtime behavior and code structure introduced by a CR.
We expect this to facilitate the comprehension of changes, especially for runtime behavior modifications, which is rather difficult by reading the source code alone.
%However, since PC is mainly done by understanding code, our design also incorporates web-based IDEs.
%With this approach, reviewers can, for example, open a preconfigured web version of Visual Studio Code with an extending software visualization from within a CR. % and do not have to checkout the code themselves.

The remainder of this paper is structured as follows.
Section~\ref{sec:design} introduces the design of our approach.
The implementation of this design is presented in Section~\ref{sec:implementation}.
Section~\ref{sec:example-application} illustrates the visual integration of static and dynamic analysis within our approach.
Scalability and limitation concerns are discussed in Section~\ref{sec:scalability-and-limitations}.
Section~\ref{sec:related-work} differentiates this work from related approaches.
Finally, Section~\ref{sec:conclusions} concludes the paper and outlines directions for future research.

%\clearpage

\begin{figure*}
	\includegraphics[width=\textwidth]{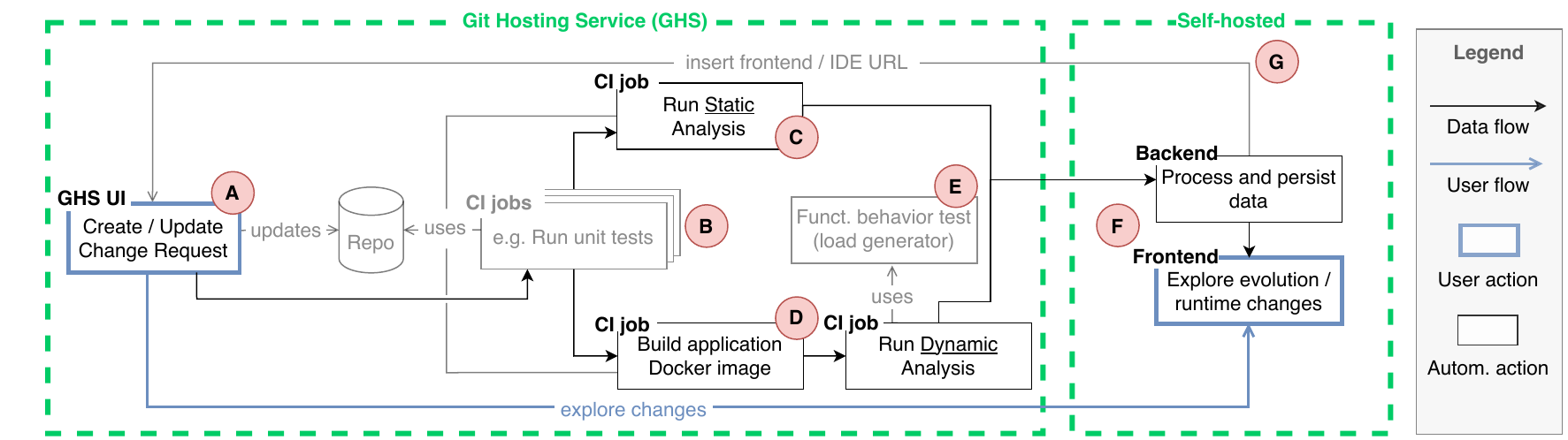}
	\caption{Conceptual data flow and architecture of our approach.}
	\label{fig:workflow}
\end{figure*}

\section{Design}\label{sec:design}
We introduce the (architectural) design of our approach by explaining its data flow illustrated in Figure~\ref{fig:workflow}.
In this context, we distinguish between actions occurring within the GHS and those executed within the self-hosted components of our approach, as indicated by the green dashed boxes.
User actions are denoted by blue-colored entities, while solid black entities represent automatically executed actions.
These automated actions operate as stateless CI jobs and as operations within the stateful backend component.
As is standard practice, the CI jobs require initial configuration and adaptation to the specific application, particularly concerning dynamic analysis, e.g., the scope of instrumentation.
%The CI jobs require an initial configuration prior to use, however, this configuration is not a recurring user action within the scope of this approach.
%As far as we know, ReviewVis (see Section~\ref{sec:related-work}) does not take the concept of CI into account.
%Instead, it uses its backend component to fetch the files of a MR.

\subsection{Initiation of the CI Pipeline}\label{sec:design-ci}
The creation of a CR and its subsequent updates (represented as new Git commits for the repository) initiate a CI pipeline execution and the corresponding data flow (Figure~\ref{fig:workflow}-A).
% This is the default behavior in professional software development.
Within the context of CI, multiple lightweight build and verification jobs must typically succeed prior to the execution of more time-intensive tasks (Figure~\ref{fig:workflow}-B).
Consequently, the static and dynamic analyses in our approach are executed in separate CI jobs and only proceed after the successful completion of the preceding jobs.

\subsection{Static Analysis}\label{sec:design-static-analysis}
The static analysis (Figure~\ref{fig:workflow}-C) produces a comprehensive snapshot that outlines the structure and the changes of the source code, based on the Git repository, i.e., the Git branch, associated with the CR.
Additionally, it computes a range of metrics for each source code file, as well as for the classes and their methods.

\subsection{Dynamic Analysis}\label{sec:design-dynamic-analysis}
For the dynamic analysis (Figure~\ref{fig:workflow}-D), it is required to instrument and execute the target software system.
Since Docker\footnote{\url{https://www.docker.com}} is the de-facto standard for deploying applications in professional software development, we run both the target application and the dynamic analysis inside Docker containers.
However, it is also necessary for the target software to perform meaningful operations during its execution.
For instance, web services are often inactive after their initial setup phase so that the instrumented code is not executed.
Developers must therefore define a load generator (Figure~\ref{fig:workflow}-E) that triggers the execution of the (desired) use cases in the target software, for example with (dockerized) JMeter\footnote{\url{https://jmeter.apache.org}} or Playwright.\footnote{\url{https://playwright.dev}}
This is similar to writing API or end-to-end tests; i.e., if they already exist, they can be reused here.
Both, the load generator and the instrumented target application are then executed within the CI job to start the recording of execution traces.

\subsection{Data Processing}\label{sec:design-data-processing}
The data resulting from the static and dynamic analyses is forwarded to the backend component, which is self-hosted outside of the GHS (Figure~\ref{fig:workflow}-F).
Within the backend, all static and dynamic analysis data is processed, persisted, and provided for later visualization.
In addition, the backend inserts a URL of the web-based frontend component into the CR description, allowing reviewers to access the corresponding visualization with a single click from within the CR.

\subsection{Visualization}\label{sec:design-visualization}
The frontend (Figure~\ref{fig:workflow}-G) combines and renders the application structure and runtime behavior in a single visualization.
At the same time, the changes introduced by the CR in both the application structure and the runtime behavior are highlighted.
This will be further explained and visualized in the upcoming Section~\ref{sec:example-application}.

%As far as we know, this differs from ReviewVis, where only the incoming structural changes are visualized, but not the rest of the application structure, let alone the dynamic runtime behavior.
%Since the code review's underlying task of PC is mainly done by understanding code in IDEs~\cite{xia2018}, we also incorporate the use of web-based development environments.
%To do this, a backend component extends the description of a CR with a URL that links to a web-based IDE for inspecting code changes.
%Provided that the IDE has an extension mechanism for the integration of web content, as is the case with most IDEs today, we can embed the visualization into the development environment.
%Code reviewers can switch the visualization on or off as required and use code proximal~\cite{Lanza2003, bassil2001, kienle2007} features, e.g., navigation from a visualized Java method call to the code or vice versa, without leaving the IDE.
%Hence, we minimize undesirable context switches between IDE and SV~\cite{sensalire2007}.
\section{Implementation}\label{sec:implementation}
We implemented the design introduced in Section~\ref{sec:design} by extending ExplorViz~\cite{VISSOFT2013,fittkau2017IST,hasselbring2020}.
ExplorViz comprises several applications, which are publicly developed on GitHub\footnote{\url{https://github.com/explorviz}} and available as Docker images.\footnote{\url{https://hub.docker.com/u/explorviz}}
Due to space constraints, we will focus on the newly developed and most important features of our approach below.
%For further implementation details~\cite{krauseglau2022ic2e,krauseglau2022vissoft,krauseglau2023IDE}, readers are kindly referred to the supplementary video that showcases the extended tool in practice.
For further implementation details readers are kindly referred to previous works~\cite{krauseglau2022ic2e,krauseglau2022vissoft,krauseglau2023IDE}.
%Section~\ref{sec:evaluation}.

%\begin{figure*}
%	\centering
%	\begin{subfigure}{.5\textwidth}		
%		\centering
%		\raisebox{20mm}
%		{\includegraphics[width=\linewidth]{images/gitlab-issue-with-explorviz-link.png}}
%		\caption{A subfigure}
%		\label{fig:sub1}
%	\end{subfigure}%
%	\begin{subfigure}{.5\textwidth}
%		\centering
%		\includegraphics[width=\linewidth]{images/visualization-timeline-composition.png}
%		\caption{A subfigure}
%		\label{fig:sub2}
%	\end{subfigure}
%	\caption{A figure with two subfigures}
%	\label{fig:test}
%\end{figure*}

%Like ReviewVis, our implementation is also designed to work for Java software.
Prior to this extension, ExplorViz only provided live-trace visualization for Java applications based on dynamic analysis, which is usually performed outside of CI environments.
To address this limitation, we implemented our design for target applications developed and built on the GitLab GHS.
Consequently, we adapted ExplorViz's dynamic analysis approach to be executed as a CI job (Figure~\ref{fig:workflow}-D).
In this context, ExplorViz utilizes NovaTec's inspectIT Ocelot (hereafter: Ocelot).\footnote{\url{https://www.inspectit.rocks}}
Ocelot is a Java agent that uses bytecode weaving to instrument and monitor Java applications.
It supports distributed tracing and offers various exporters, including exporters for the industry standard OpenTelemetry,\footnote{\url{https://opentelemetry.io}} which is used by ExplorViz.
To realize our new design for the dynamic analysis, we employ a dockerized version of Ocelot in CI.
Additionally, we leverage GitLab CI environment variables to incorporate the hash (as a unique identifier) of the current Git commit into each execution trace recorded by Ocelot.
This approach enables us to later find the correct instances of runtime behavior for a given Git commit.
Consequently, it allows us to map runtime behavior to structural evolution data, the latter being the outcome of the static code analysis.

To realize the static code analysis outlined in our design (Figure~\ref{fig:workflow}-C), we developed a static code analyzer for Java source code.
This tool, referred to as the \emph{code agent}, is written in Java and can be used in or outside of CI environments.
When executed in a CI job for a Git commit, the code agent creates a list of all Java classes that are contained in the Git branch associated with the relevant CR.
Additionally, it analyzes the changed Java classes of the Git commit via the JavaParser library.\footnote{\url{https://javaparser.org}}
%Specifically, JavaParser creates an abstract syntax tree that is traversed and processed for each source code file.
As a result, the static analysis includes, for example, the recognition of (abstract) classes, inheritance, interfaces, and method declarations.

\begin{figure}
	\includegraphics[width=\linewidth]{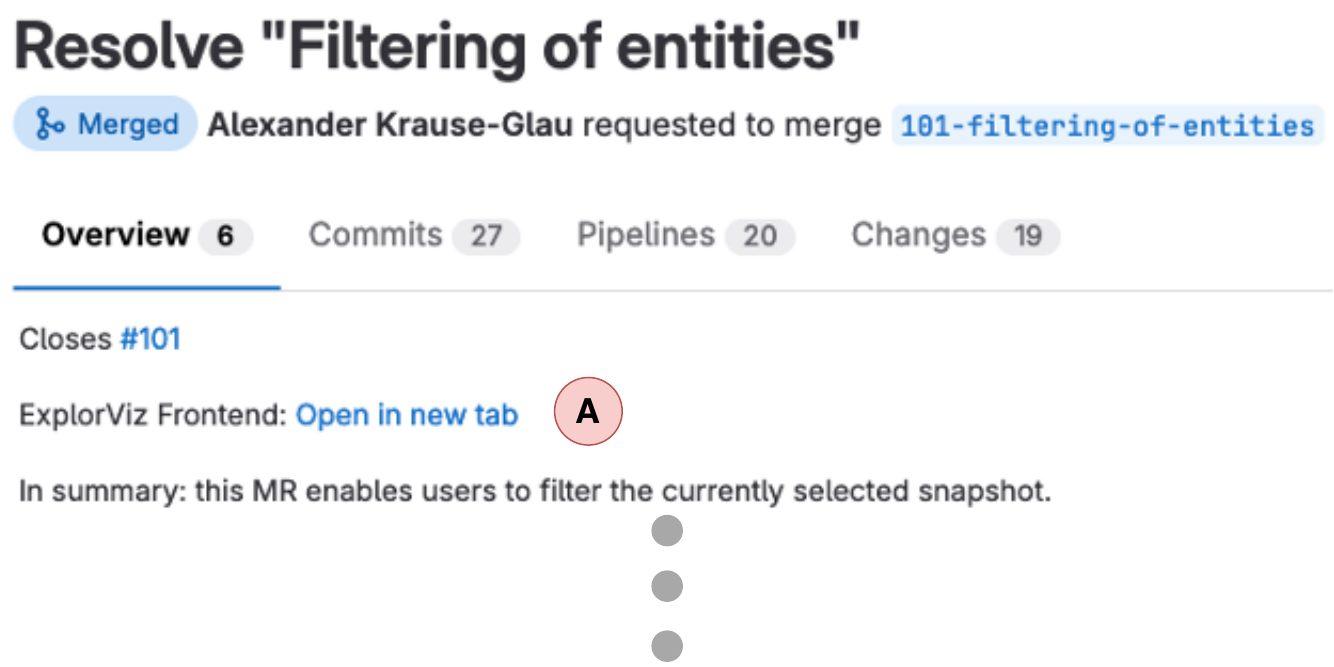}
	\caption{Excerpt of an example GitLab issue. Code reviewers can click on the ExplorViz frontend link to access the related visualization.}
	\label{fig:issue}
\end{figure}

Both, the static code analysis data and the runtime data are then sent to the ExplorViz backend (Figure~\ref{fig:workflow}-F) via gRPC,\footnote{\url{https://grpc.io}} which completes the CI jobs.
Here, the code service and the span service receive the static analysis data and the runtime data, respectively.
Each service processes, persists, and provides the prepared visualization data via an HTTP API.
Storing the data allows users to select / compare previously analyzed Git commits to examine, for example, the evolution of a software system in terms of structural and behavioral changes.
The commit comparison described here is related to other SV approaches that address code evolution~\cite{ardigo2022M3tricity,ciani2015UrbanIt,occhipinti2023Syn}.
%By processing we mean the calculation of the changes, both for the structural development and for the runtime behavior.
Eventually, all data will be made available to the extended ExplorViz frontend component, a web-based WebGL application~\cite{krauseglau2022ist}.
Here, users can select analyzed applications and explore the changes introduced by the CR.
In the upcoming Section~\ref{sec:example-application}, we will demonstrate this process by exploring the software city visualization of our approach using an example software system.

\begin{figure}
	\includegraphics[width=\linewidth]{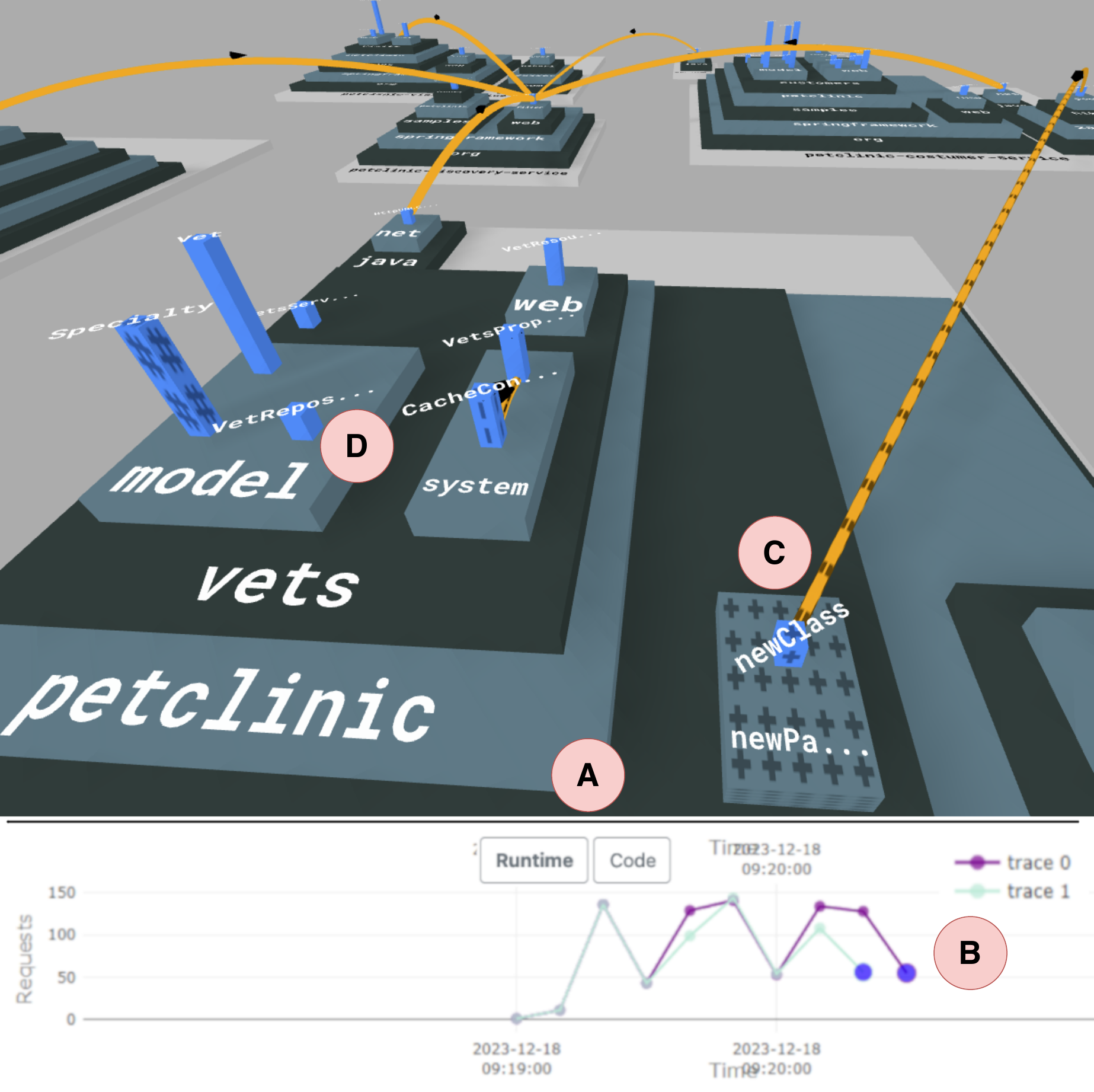}
	\caption{Software cities visualize the distributed Spring PetClinic. In this color scheme, the packages are light and dark gray. The height of the blue buildings (classes) indicates the current number of instances for a class based on the runtime behavior. Orange pipes show method calls captured by dynamic analysis. Textures, e.g. plus signs, mark changes between the selected Git commits and their runtime snapshots.}
	\label{fig:codecity}
\end{figure}

\section{Visual Integration of Static and Dynamic Software Analysis}\label{sec:example-application}
In this section, we present the visual integration of static and dynamic analysis data in our approach.
To illustrate this, we introduce ExplorViz's extended software city visualization using an example software system.
%Due to space constraints, we kindly refer readers to the supplementary video, which showcases the complete implementation of our design illustrated in Figure~\ref{fig:workflow}, including the GitLab integration.

\begin{figure*}[htbp]
	\centering
	\begin{minipage}[b]{0.3\textwidth}
		\centering
		\includegraphics[width=\textwidth]{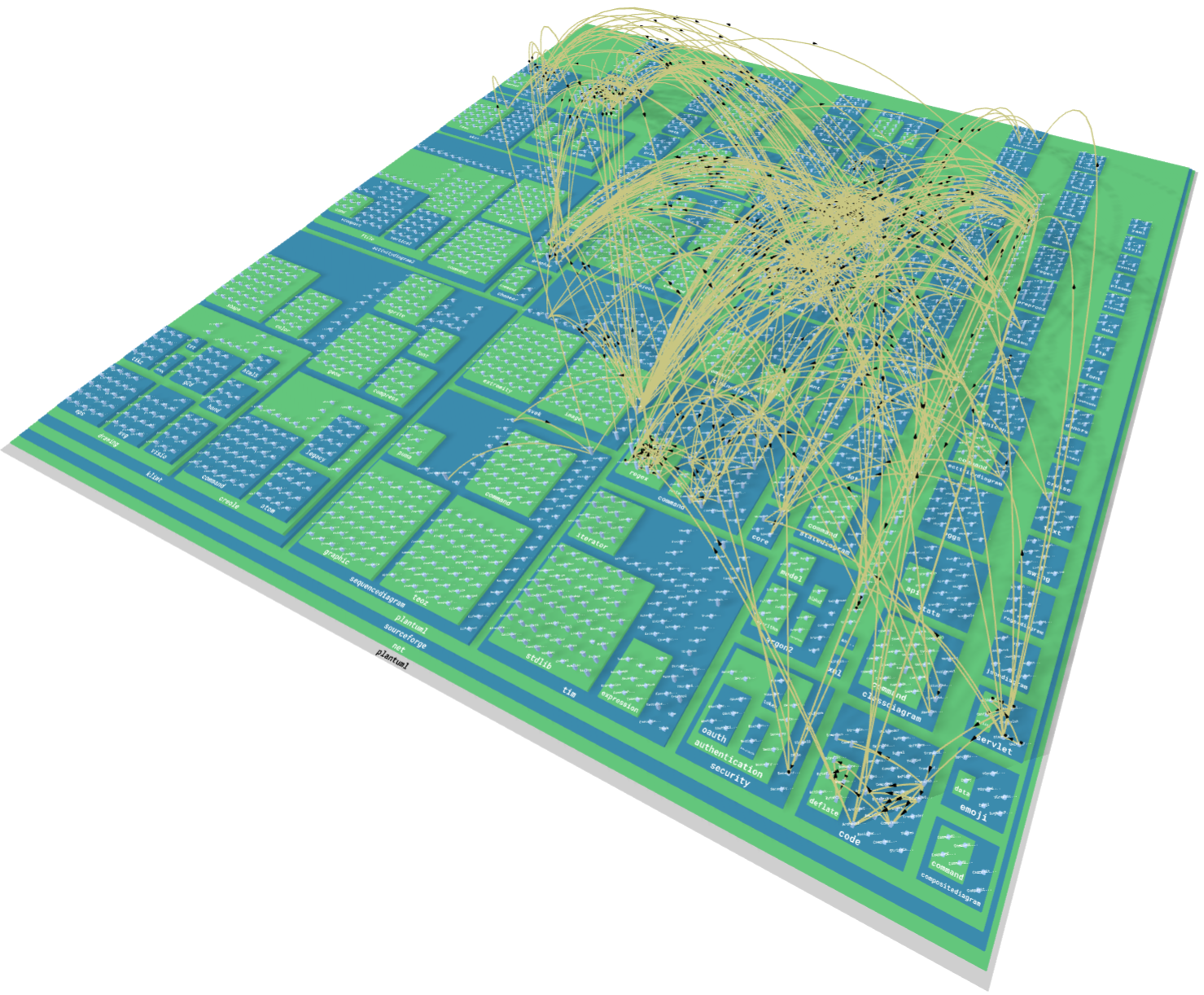}
		\subcaption{Unfiltered comparison showing all commit-related data. Four data sets are dynamically combined for this visualization.}
		\label{fig:image1}
	\end{minipage}
	\hfill
	\begin{minipage}[b]{0.3\textwidth}
		\centering
		\includegraphics[width=\textwidth]{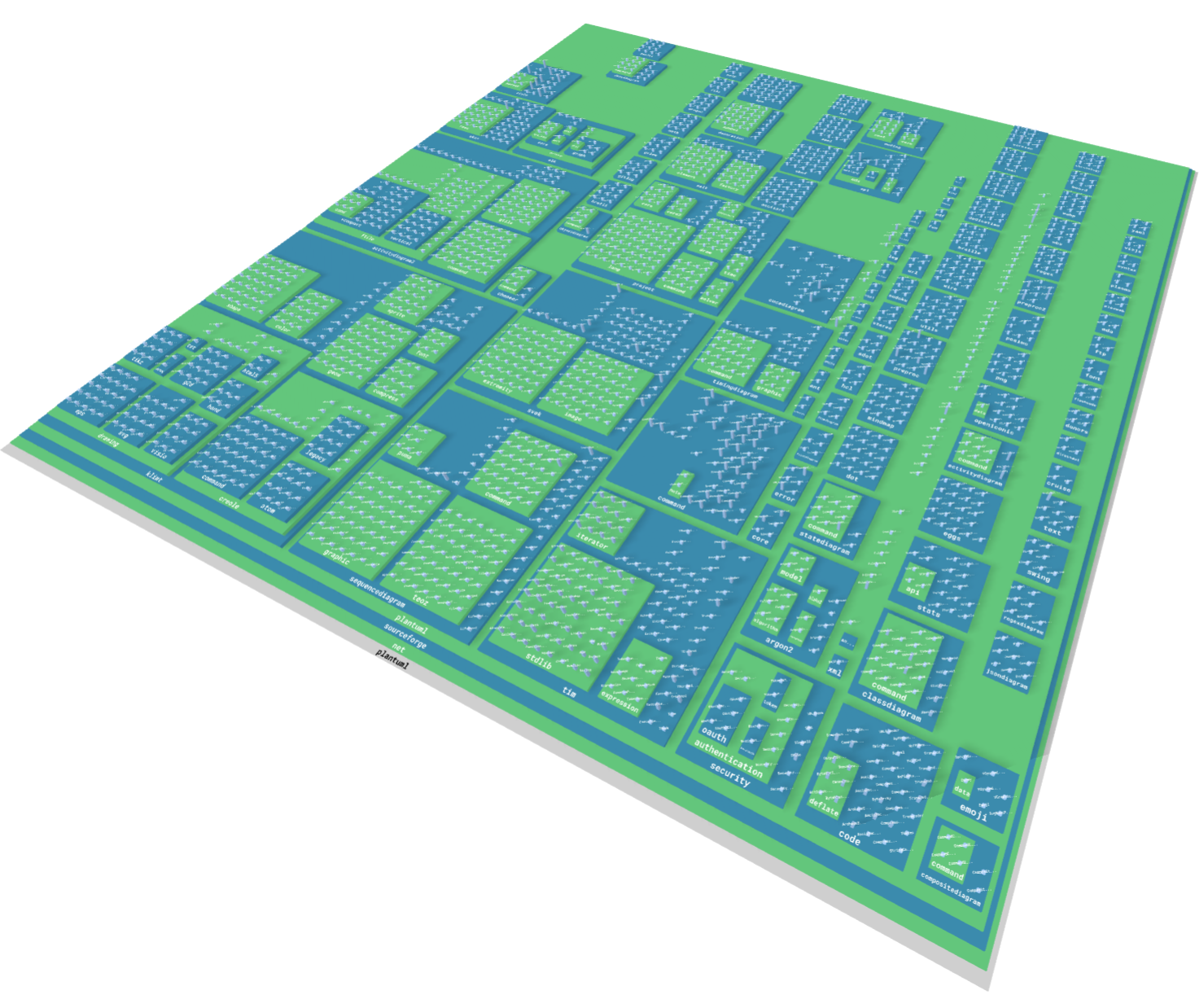}
		\subcaption{Filtered comparison visualizing the compared static analysis data without any runtime information.}
		\label{fig:image2}
	\end{minipage}
	\hfill
	\begin{minipage}[b]{0.3\textwidth}
		\centering
		\includegraphics[width=\textwidth]{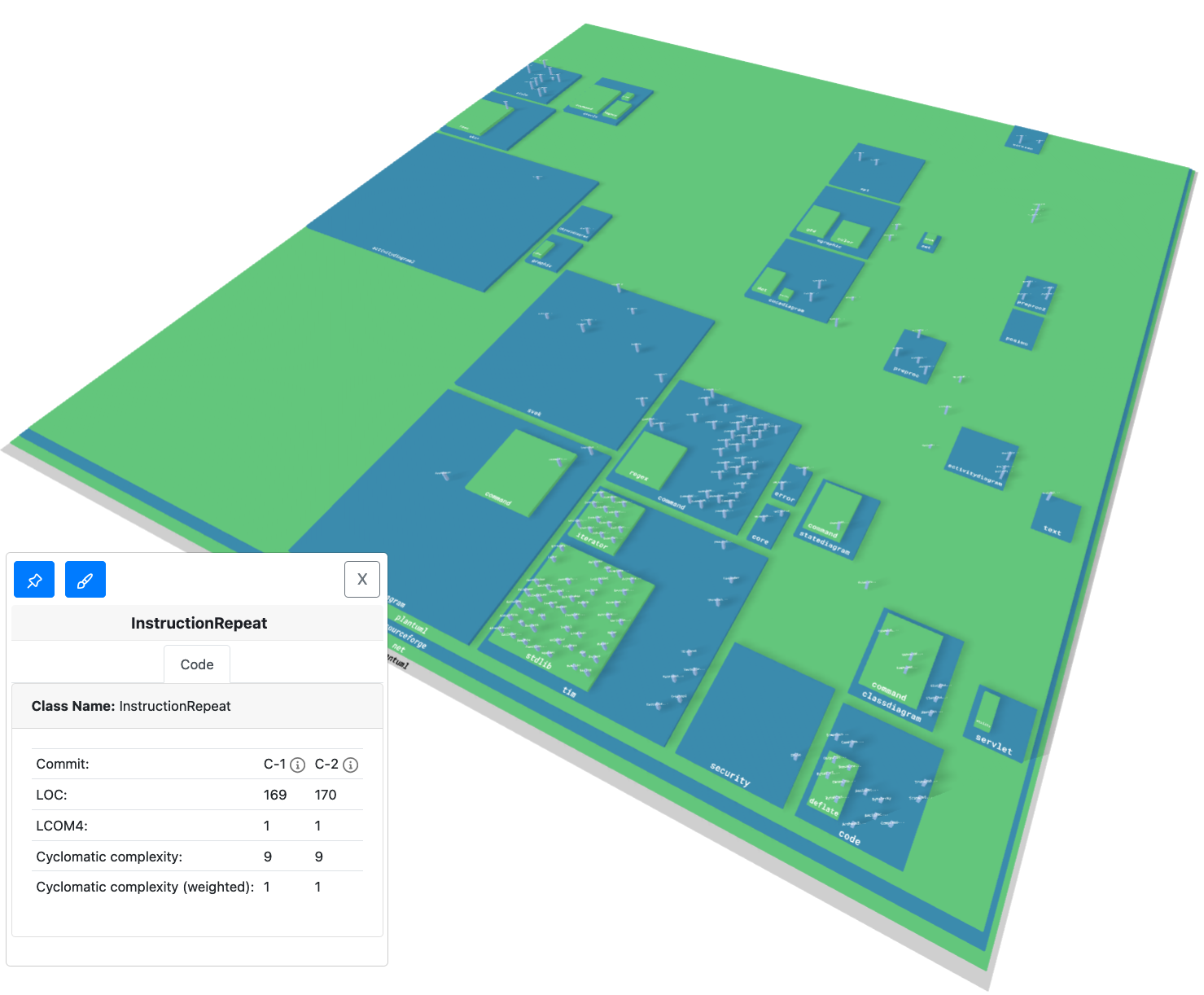}
		\subcaption{Filtered comparison emphasizing only the code differences. Comparison details are displayed upon hovering over an entity.}%as determined by the processing of the static analysis data.}
		\label{fig:image4}
	\end{minipage}
	\caption{Software city visualizations comparing two commits of PlantUML at varying levels of detail. Users can select a combination of analysis data types to be visualized using toggle buttons.}
	\label{fig:software-city-scalability}
\end{figure*}

The extended ExplorViz frontend is accessible via a web browser, either directly or via a link from within a CR as illustrated by Figure~\ref{fig:issue}-A.
By utilizing query parameters, users can specify which particular Git commit or which comparison of Git commits of an analyzed application should be visualized.
By default, applications that are not configured via a query parameter are rendered using their last known Git commit.
This feature is utilized by the extended ExplorViz backend when updating the CR description. 
As a result, reviewers are presented with the correct visualization when clicking the ExplorViz frontend link in the CR description (Figure~\ref{fig:workflow}-G).

Figure~\ref{fig:codecity} depicts the software city visualization of our approach.
ExplorViz currently visualizes all data using the city metaphor, as this has a positive effect in the context of program understanding~\cite{wettel2011ControlledExperiment,galperin2022}.
For the sample application, we use the distributed version of the Spring PetClinic (hereafter PetClinic).\footnote{\url{https://github.com/spring-petclinic/spring-petclinic-microservices}}
The PetClinic consists of multiple microservices and a frontend component that communicates with all services depending on the executed use case, such as booking a visit to the veterinarian for a user's pet.
The accompanying screenshot illustrates the visualization of both the runtime behavior and code structure for the PetClinic.
In this context, multiple software cities, i.e., one for each service of the PetClinic, are visualized simultaneously.
Additionally, a specific Git commit comparison for the service in the foreground has already been selected.
%This can be changed at any time via the code tab (Figure~\ref{fig:codecity}-A) as demonstrated in our supplementary video.

This can be changed at any time via the code tab (Figure~\ref{fig:codecity}-A).
In the example shown Figure~\ref{fig:codecity}, the two compared Git commits also have associated runtime behavior for the encompassing service, as illustrated by the runtime tab.
Two runtime snapshots, i.e., aggregated runtime behavior over ten seconds, are selected and compared (Figure~\ref{fig:codecity}-B).
For this comparison, we use textures because colors are already associated with other meanings in our visualization approach.
Figure~\ref{fig:codecity}-C depicts a new Java package with a new class, indicated by a plus sign texture.
The new entities were identified based on the comparison of the code structure between the two selected Git commits.
Additionally, Figure~\ref{fig:codecity}-C shows a new communication line, using the same texture, derived from the comparison of the runtime behavior for the selected runtime snapshots.
This indicates a new method call during the runtime behavior which could be caused by the proposed code changes of the CR.
Removed or modified entities are represented by other textures, as illustrated in Figure~\ref{fig:codecity}-D, while hovering with a mouse over an entity displays additional information.

\section{Scalability \& Limitations}\label{sec:scalability-and-limitations}
In the context of data visualization, an excess of information within a single visualization can impede comprehension.
Therefore, tools such as SV must account for the scalability of information~\cite{kienle2007}.
Prior to this work, ExplorViz already allowed users to interactively filter entities in software cities.
This feature allows users to more easily locate specific entities, such as classes with a particular instance count, thereby enhancing the users' ability to understand complex software structures.
With the introduced extension of ExplorViz in this work, we expand the filtering feature.

Figure~\ref{fig:software-city-scalability} depicts the same software city visualization comparing two commits of PlantUML\footnote{\url{https://plantuml.com}} with selected runtime snapshots (see Section~\ref{sec:example-application}) at varying levels of detail.
In such situations, the ExplorViz frontend, by default, combines all four data sets, i.e., processed code and runtime data for each of the two commits, into a single visualization.
The resulting unfiltered comparison can be seen in Figure~\ref{fig:image1}.
Although applied textures indicate if classes and packages have been added, deleted, or modified, these changes are not easily visible in the unfiltered view.
To address this, the visualization can be dynamically adjusted using a set of toggle buttons that allow users to select the type of data to be visualized.
For instance, Figure~\ref{fig:image2} displays a filtered comparison where the software city visualization includes only the combined code structures of the two commits. In this view, certain packages and classes are absent as they were present only in the runtime behavior.
Differences in packages and classes are still indicated with textures.
Furthermore, users have the option to visualize only the differences between the two commits.
This is demonstrated in Figure~\ref{fig:image4}, which shows the actual code differences in the visualization.
Detailed comparison information, such as the lines of code for a class between the two commits, is displayed upon hovering over an entity.

Beyond information scalability, rendering scalability, defined as the capability to scale the visualization performance to manage increasing amounts of data, is a critical requirement for SV tools~\cite{kienle2007}.
This is one of the current limitations of ExplorViz.
Although users can dynamically change the visualized commits and related runtime snapshots in the frontend, these actions currently trigger on-demand processing of the desired data without employing any reduction techniques.
Consequently, visualizations for larger datasets, such as the PlantUML example depicted in Figure~\ref{fig:software-city-scalability}, experience significant delays when updating upon the selection of different commits or runtime snapshots.

Another limitation is the current capability of the code agent (see Section~\ref{sec:implementation}) to resolve Java import paths in analyzed class files.
For instance, the code agent successfully identified over 2,400 Java classes. However, the logs of the code agent indicated that classes were skipped due to incomplete analysis of their source code.

% Big Data, long processing on demand
% An example for that is the visualization depicted in Figure x.
% Here, we see a software city showing the comparison between two commits of PlantUML.
% Over 2400 classes, but only a few hundred show due to import resolution problems.
% Also layout is not stable when commits or runtime is changed.

\section{Related Work \& Novelty}\label{sec:related-work}
The combination of static and dynamic analysis has a wide range of applications~\cite{ HeuzerothCombineForInteractionPattermns, fan2007DesignPatternDetectionViaCombination,sellami2022CombineForDecomposition,thair2015CombineForUnitTestDistribution}.
% nach Annahme: krause2020
However, in the context of modern code review, related SV approaches generally only use static code analysis~\cite{balci2021AugmentingCodeReviews, gasparini2021ChangeViz, tymchuk2015CodeReviewVeniTool}.
Dynamic analysis, on the other hand, is often used by code coverage tools for CRs~\cite{oosterwaal2016VisualizeCoverageForCodeReview}\footnote{\url{https://cobertura.github.io/cobertura}} or SV approaches that are not concerned about the review process~\cite{dashuber2022ControlledExperimentJournal}.
Below we present the work that we consider to be the most related to ours in terms of system integration, perceived degree of maturity, and using SV to facilitate the review task.

In 2022, Fregnan et al. introduced ReviewVis~\cite{fregnan2023GraphVisualizationMergeRequestCodeReview}, a tool aimed at facilitating the comprehension of Java code changes introduced by CRs.
ReviewVis enhances GitLab's merge request (MR) interface with interactive 2D force-directed graphs.
These graphs feature nodes representing classes, interfaces, methods, or non-Java files, colored to indicate their change status (e.g., added, deleted, or changed).
A backend component performs static analysis on the changed source code, creating and comparing abstract syntax trees for both the source and target branches to determine the change status of each file.
The visualization is displayed via a Google Chrome extension, which allows users to navigate between the graph and the related code in the GitLab MR view.
Surveys with professional developers indicate that ReviewVis aids in comprehending and navigating code changes, though it offers limited help for small MRs and requires improvements for larger MRs.
Although ReviewVis and our approach share the goal of facilitating code review through SV, they differ in several design aspects.
Our approach integrates both static and dynamic analysis within software cities, whereas ReviewVis relies solely on static analysis.
Additionally, our use of the city metaphor contrasts with ReviewVis's graph rendering using D3.js\footnote{\url{https://d3js.org}}.
The authors of ReviewVis also consider the potential application of the city metaphor for large MRs based on participant feedback.

Recently, Augustinowski et al. presented their ongoing study to determine the necessary features for their SV tool SEE to enhance the code review process~\cite{augustinowski2024SEECodeReview}.
SEE and ExplorViz share concepts such as the SV metaphor and support for collaborative use~\cite{krauseglau2022vissoft}.
Currently, SEE is in an early development stage, with potential CI integration concepts that are comparable to our approach not yet formulated.
At this stage, SEE includes an in-tool web browser, likely for CR description access, whereas ExplorViz automatically links to the corresponding CR visualization in the description.

%For reasons of space, we cannot go into all the differences between ReviewVis and our work.
%In the following, however, we will describe the event flow of our apparoach, followed by implementation details of our currently realized prototype, and highlight notable similarities and distinctions between the two works.
\section{Conclusions}\label{sec:conclusions}
In this paper, we report on the novel and in-progress design and implementation of a web-based approach capable of combining static and dynamic analysis data in software city visualizations.
In comparison to related works, our approach incorporates the collection and visualization of both (changed) runtime behavior and code structure introduced by a pull / merge request.
Our architectural tool design incorporates modern web technologies such as the integration into common Git hosting services.
As a result, code reviewers can see how the modified software evolves and executes its use cases, which is especially helpful for the underlying task of program comprehension.
We have presented the implementation of our design through the extension of ExplorViz~\cite{VISSOFT2013,fittkau2017IST,hasselbring2020}, which so far only used dynamic analysis data.
The resulting source code can be found on GitHub.\footnote{\url{https://github.com/explorviz}}
Additionally, we presented and discussed ExplorViz's extended software city visualization using an example software system.
We provide a video of our approach online, which showcases the implementation of our design.

We expect that the visual integration of both static and dynamic software analysis facilitates the task of code review, particularly within the context of distributed software systems.
We plan to investigate this assumption in future research, subsequent to addressing current limitations of our approach.
Additionally, we aim to support the new code review mode in ExplorViz's collaborative~\cite{krauseglau2022vissoft} and code-proximal approaches~\cite{krauseglau2023IDE}.
For instance, ExplorViz's Visual Studio Code extension could automatically leverage the code editor's internal Git feature.
Consequently, users would be presented not only with the comparison visualization but also with the associated code changes in a predefined web-based development environment.

\section*{Acknowledgment}
The authors would like to thank Julian Pleines, Lennart Ideler, and Arash Giv for their contributions with implementing and evaluating some of the features presented in this paper.

\clearpage
\providecommand{\doi}[1]{DOI: \href{https://doi.org/#1}{#1}}
\bibliographystyle{myIEEEtran}
%\interlinepenalty=10000
%\balance
\bibliography{explorviz-vissoft-2024-combination}

\end{document}